\documentstyle[11pt,newpasp,twoside,epsfig]{article}
\def\edcomment#1{\iffalse\marginpar{\raggedright\sl#1\/}\else\relax\fi}
\reversemarginpar

\begin{document}
\title{Cosmic Shear with Keck: Systematic Effects}
\author{Richard Massey, David Bacon, Alexandre Refregier}
\affil{Institute of Astronomy, Madingley Road, Cambridge CB3 0HA, UK}
\author{Richard Ellis}
\affil{California Institute of Technology, Pasadena CA 91125, USA}

%
%

\begin{abstract} Cosmic shear probes the distribution of dark matter
via gravitational lensing of distant, background galaxies. We
describe our cosmic shear survey consisting of deep blank fields
observed with the Keck II telescope. We have found biases in the
standard weak lensing analysis, which are enhanced by the elongated
geometry of the Keck fields. We show how these biases can be
diagnosed and corrected by masking edges and chip defects.
\end{abstract}

%
%


\noindent
Images of background galaxies become weakly distorted by
gravitational lensing as light travels through intervening
large-scale structure to the observer. These distortions provide a
direct measurement of the mass distribution in the universe, without
assumptions about the nature, state or luminosity of the dark
matter.

This ``cosmic shear'' effect is small, with the axis ratio of a
typical galaxy being elongated by only about 1\%. However, adjacent
lines of sight have passed through the same large-scale structures
and close ($\sim$1') pairs of background galaxies receive coherent
distortions. After removing systematic shape measurement biases, we
may average over many galaxy pairs to detect this signal. We present
a survey covering 0.6 square degrees to $R\sim26$, scattered
randomly around the sky in 173 pointings of the Echelle Spectrograph
and Imager on Keck II. After cuts, our final catalogue contains
63,000 galaxies. 


%
%

\vspace{6mm} \noindent
{\bf Shear Measurement and Elimination of Systematic Biases}
\vspace{3mm}

\noindent
Measurement of such a small effect demands careful data
reduction and elimination of systematic errors. Bias subtraction and
flat fielding were optimised using the background level and the
overscan regions at the edge of individual exposures. Three slightly
dithered exposures were then astrometrically realigned and co-added
to produce each survey field. As a by-product, this allowed the
monitoring of instrumental distortion at different telescope
positions. Keck proves ideal for cosmic shear, with negligible
$<$0.3\% internal distortions even at the edges of the 2'$\times$8'
field of view. The PSF shape was measured from stars across each
reduced field.  The isotropic and anisotropic contributions of the
PSF to galaxy shapes was then corrected using the {\tt KSB} method
(Kaiser, Squires, \& Broadhurst 1995), which is well tested and
calibrated (Bacon et al. 2001; Erben et al. 2001).


\begin{figure}  \epsfig{figure=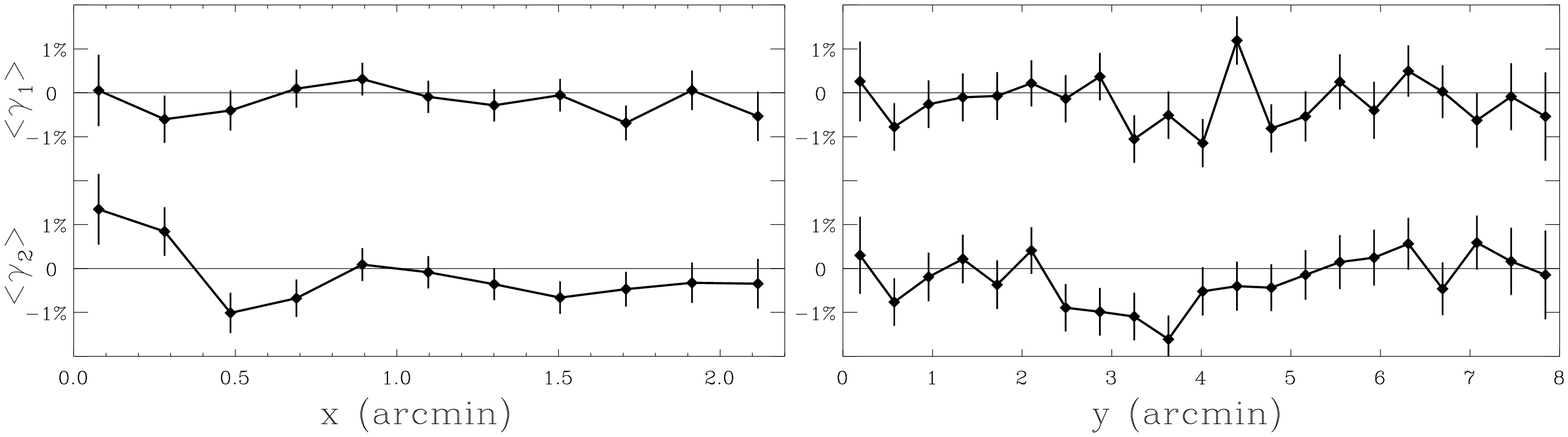,width=133mm}
\caption{Mean shear of all galaxies in our Keck fields as a function
of their position on the CCD, after the corrections described in the
text. This should be consistent with zero in the absence of
systematics.  Overall, we find $\langle\gamma_1\rangle = -0.02\% \pm
0.16$ and $\langle\gamma_2\rangle = -0.29\% \pm 0.16$.} \end{figure}

However, the long-thin geometry of the ESI field enhanced two
unforeseen biases in this standard reduction procedure. Firstly, the
PSF interpolation in the short $x$ direction was often overfit by
{\tt KSB}'s simple polynomial fitting. The divergent PSF model
overcompensated for the true smearing and spuriously elongated in
the $y$ direction galaxies near to the sides. This effect revealed
itself as a residual shear offset (similar to that in CFHT data of
van Waerbeke et al. 2001) and in the final shear correlation
functions as a $\sim$1.1\% shear excess in pair separations around
2', the width of the chip. A more adaptative algorithm was written,
which found the smallest degree polynomial possible for a
significant fit and which could be monitored to remove any anomalous
stars by hand.

Edge effects (including not only the boundaries of the CCD but also
chip defects and saturated stars) were also of concern for such a
narrow field of view. Galaxies near an edge on any one of the three
dithered exposures are cut in half and appear aligned to that
boundary. Even if not exactly on the boundary, the flat fielding was
poorer near edges ($\sim$10$^{-4}$ gradient) and image co-addition
failed because of differing background levels in the dithers. If
strips of galaxies are not excluded from the final catalogue, the
overall mean shear increases by $\sim$2\% or $\sim$1\% respectively.
With masking and PSF fitting as described, this offset disappears.
Figure 1, a useful diagnostic tool, shows our final results.

%
%

\vspace{6mm} \noindent
{\bf Conclusions}
\vspace{3mm}

\noindent
Two previously unforeseen biases in the standard weak
lensing procedure, enhanced by the elongated Keck field geometry,
were isolated and removed by CCD masking. Further treatment of
systematics, results, and their cosmological implications will be
presented in a forthcoming paper (Bacon et al. 2002, in
preparation).

\end{document}